\documentstyle[preprint,revtex]{aps}
\draft
\tightenlines
\begin{document}

\begin{title}
SKYRMIONS AND THE NUCLEAR FORCE
\end{title}

\vspace{0.5cm}

\author{Niels R. Walet and  R. D. Amado}
\begin{instit}
Department of Physics, University of Pennsylvania,
Philadelphia PA 19104-6396
\end{instit}

\begin{abstract}
The derivation of the nucleon-nucleon force from the
Skyrme model is reexamined. Starting from previous results for the
potential energy of quasistatic solutions,
we show that a calculation using
the Born-Oppenheimer approximation
properly taking into account the mixing of
nucleon resonances, leads to substantial central
attraction. We obtain a potential that is in
qualitative agreement with phenomenological potentials.
We also study the non-adiabatic corrections, such as the
velocity dependent transition potentials, and
discuss their importance.
\end{abstract}
\section{Introduction}
     A central problem of strong interaction theory is to derive
the nucleon-nucleon interaction from more fundamental
principles.  It is well established that quantum chromodynamics (QCD) is the
correct theory of strong interactions.
The phenomenology of the nucleon-nucleon interaction suggests
a description in terms of nucleons and the exchange of mesons, while
QCD is a non-abelian gauge theory of quarks of three colors
interacting by exchange of gluons.
These two pictures, one of quarks and
gluons and the other of nucleons exchanging mesons are difficult to
reconcile.  The notions of asymptotic freedom and its counterpart
confinement suggest that quarks and gluons are the appropriate degrees
of freedom for short distances and high energy, but are not for the
large distances and relatively low energies of the nucleon interaction,
where the quark picture becomes strongly non-perturbative.
The problem is to make a theory, starting from QCD, that obtains the
nucleons and their meson exchange.  If such a theory agrees with
the vast body of nucleon-nucleon interaction
phenomenology, it will have, in some
sense, ``derived'' traditional nuclear physics from QCD.  This paper
reports the details of an effort based on the Skyrme model \cite{Skyrme,%
SkyrmRefs} to make that connection.  A brief
summary of these results have been presented elsewhere \cite{PRL}.

 The Skyrme model describes a nonlinear, classical field theory
of interacting pions.  This seems far from QCD, but in fact the
Skyrme Lagrangian has been shown to be
an effective theory of QCD
in the large distance (long wave length) non-perturbative regime in
the limit of a large number of colors ($N_{\rm C}$). There is no unique path
from such a classical theory back to quantum chromodynamics with three
colors, but attempts to construct alternate classical pictures have
shown that there is surprisingly little leeway \cite{JacksonHighOrder},
and work to extract nucleon phenomenology from the Skyrme Lagrangian
has been relatively successful with results that are fairly
insensitive to  variations in details of the theory.  Hence the Skyrme
Lagrangian provides a robust, QCD-based starting point for studies
of nucleon and nucleon-nucleon properties in the long wavelength,
non-perturbative regime.

     The Skyrme Lagrangian possesses a topologically conserved charge,
$B$, that Skyrme \cite{Skyrme}
identified with baryon number. The $B=1$ solutions
of the Skyrme Lagrangian yield an appealing  picture of the
nucleon and its excited states, but since the theory is a large
$N_{\rm C}$ approximation, corrections for real nucleon observables are
expected to be of order $1/N_{\rm C}$ or $\sim 30 \%$.  An interesting
and important problem area is the calculation of these $1/N_{\rm C}$
corrections.  These are first quantum corrections, and their
calculation faces the standard uncertainties found in going
from any classical theory, particularly a non-linear field theory,
to a quantum theory.

     The relative success of the Skyrme picture for the $B=1$ system
invites its extension to $B=2$ and in particular to the problem
of the nucleon-nucleon interaction.  The phenomenology of the $NN$
interaction suggests that, with the possible exception of the repulsive
core, the interaction is well
described in terms of multiple pion exchange.
Thus the Skyrme model which expresses the physics in terms of
interacting pions, seems to be a good starting point.  Skyrme himself
\cite{Skyrme}
noted that for very large separation, the $NN$ interaction reduces
to one pion exchange in his model.  The key question then is whether
the model can also give the central mid-range attraction that is
responsible for nuclear binding.  There are three problems
that must be overcome in addressing this question.  The first is
the solution of the classical $B=2$ field equations to obtain
and describe $B=2$ configurations.  The second is the problem of
obtaining  nucleon-nucleon interactions
from Skyrmion-Skyrmion interactions.
The third is the introduction of $1/N_{\rm C}$ corrections.  As we shall see
the last two are inexorably mixed.  The need for careful $1/N_{\rm C}$
corrections is much greater in the $B=2$ problem than for $B=1$, because
the only way presently known to calculate the interaction energy
is to calculate the energy of the total $B=2$ system and then
subtract the energy of two free $B=1$ Skyrmions. This is a difficult,
delicate, and dangerous process since the interaction energy is a
very small fraction of the total.  In this paper we present
a procedure for carrying out the three steps.  Our
procedure yields a $NN$ interaction closely resembling
standard phenomenology.  The interaction has a
one pion exchange tail at long distance, a short range repulsion,
a tensor force and, best of all, a strong, mid-range
central attraction--all in substantive agreement with phenomenology.

     The problem of solving or even of describing the
$B=2$ interacting Skyrmion system is complex.
There are only two known stable (time-independent)
$B=2$ configurations.  One is the trivial one of two infinitely
separated $B=1$ Skyrmions, and the other is a compact bound state of
toroidal baryon density in which the identity of the two individual
Skyrmions is lost.  To obtain an interaction energy one must define
a set of intermediate, unstable, static Skyrmion configurations at
finite separation.  Each configuration defines a point on an adiabatic,
six dimensional, collective manifold, called the unstable
manifold by Manton \cite{MantonPRL}.  Each point on the manifold is
stabilized
 by imposing a constraint.  The dimensionality of this manifold makes
it virtually
impossible to calculate the interaction energy at every point.
Fortunately much of the interaction energy can be described by
studying only a very small part of that manifold.  But on that
sub-manifold the description must be accurate.  As we emphasized
above, the interaction energy is calculated as the difference between
large numbers. Hence seemingly small inaccuracies in configurations
can lead to large discrepancies in interaction energy.  It is this
problem that has plagued early attempts to study the $NN$ interaction
in the Skyrme model with the product ansatz.  This ansatz, also first
given by Skyrme \cite{Skyrme}, describes the $B=2$ system as the
product of two
$B=1$ systems.  The configurations so obtained are not solutions of the
constrained non-linear field equations, and hence the energy they
yield is above the true energy.  For very large separation these
differences are not serious, but for intermediate and short distances
they are fatal.
  For example the product ansatz cannot describe the
toroidal lowest $B=2$ state.  Early efforts \cite{Jackson,VinhMau}
to study the $NN$ system
with the product ansatz developed much of the vocabulary and sense
of the problem, but in the end failed to give any mid-range attraction.
 Surprisingly enough one finds a repulsive core in this
ansatz, but it has been shown that this does not have the correct g-parity
\cite{gparity}.
A far better approach to the $B=2$ system
is direct numerical solution of the constrained,
non-linear, field equations.  Recently Walhout and Wambach (WW) \cite{WW}
have
produced just such solutions.  They form the starting point of our work.

How does one go from the Skyrmion-Skyrmion  interaction
energy calculated on the interesting sub-manifold of the unstable
manifold to the $NN$ interaction?  A commonly used prescription
is to project the Skyrmions to nucleons at each separation.
This method has been used extensively in the early product ansatz
work.  Using the same approach WW find weak central attraction.
The nucleons only requirement is too strong, however. Phenomenologically
all that is required is that the
particles be nucleons asymptotically.  When they approach and interact
they can deform and be excited, for instance to the $\Delta$ state,
as the dynamics dictates.
This state mixing must be very important, since in the solutions
for the interacting Skyrmions found by WW, the
individual Skyrmions deform strongly as they interact, finally loosing
their
individuality completely in the toroidal state.
If the time scales (or equivalently energy scales) for the radial motion
and
the excitation of Deltas separate, the
contribution of this distortion to the energy can be calculated
in the Born-Oppenheimer approximation.  Finite $N_{\rm C}$ effects are
critical in  making the time scales appropriately distinct.  It is this
combination of finite $N_{\rm C}$ effects and Born-Oppenheimer that is
crucial to obtaining the phenomenological potential.  Each idea has
been used before \cite{corrections},  but it is
the combination of exact solution of the classical field
equations, state mixing using the Born-Oppenheimer
approximation and finite $N_{\rm C}$ effects that together
makes physical sense and gives the phenomenologically correct
potential as we shall see below. We shall first treat the effects of the
distortions perturbatively, where it is clear that the distortion makes
an attractive contribution. Since the contribution of the distortions
becomes
large at small distances, we turn to diagonalization of the
 Born-Oppenheimer Hamiltonian to check the perturbative results.
Once we make the Born-Oppenheimer separation we can calculate explicit
corrections to the adiabatic picture. These take the from of transition
potentials between the different adiabatic channels, as well as some
velocity
dependent potentials similar to Berry's phase terms.

    In Section II we set out some key notions.  We review the Skyrme
Lagrangian, the adiabatic manifold for interacting $B=2$ Skyrmion
systems, review the algebraic method for projecting baryon states
from Skyrmions and for introducing finite $N_{\rm C}$ effects, and show
how these algebraic methods permit a description of the Skyrmion-Skyrmion
interaction on the unstable manifold.  This section brings together
the known tools we need for the problem.  Section III presents
our derivation of the $NN$ interaction from the Skyrmion interaction
for finite $N_{\rm C}$.  It gives the adiabatic interaction derived
first in perturbation theory and then using the Born-Oppenheimer
approximation.
The formalism for obtaining the transition potentials is also given here.
Our results are presented in Section IV.  As
we have discussed above, the potentials are close to static $NN$ potentials
obtained from the data.  Section V gives a brief summary and conclusions.
Some of the algebraic details of the perturbation
treatment are presented in the appendix.

\section{Some key notions}
\subsection{The Skyrme model}
The Skyrme model is a non-linear field theory that can be realized in
terms of an SU(2)-valued matrix field $U$, with Lagrangian density
\begin{equation}
{\cal L} = \frac{f_\pi^2}{4} {\rm Tr}[
\partial_\mu U(x) \partial^\mu U^\dagger(x)]
+ \frac{1}{32g^2}{\rm Tr}[ U^\dagger\partial_\mu U,
U^\dagger\partial_\nu U]^2.
\end{equation}
The model is covariant, as well as invariant under global
SU(2)-rotations that are identified with the isospin symmetry. As was
discovered by Skyrme the model has a topologically conserved quantum number,
which is identified as the baryon number $B$.
The $U$ field is interpreted as a combination of a scalar $\sigma$ field
and an isovector pion  field, $U=\sigma + i\vec{\tau} \cdot \vec{\pi}$.
The $\sigma$ field is not an independt physical field due to the unitarity
constraint on $U$.

The standard time-independent solution to the classical field equations
for $B=1$ is the defensive hedgehog, where the pion field points radially
outward,
\begin{equation}
U_1(\vec{r}) = \exp( i\vec{\tau} \cdot \hat{r} f(r) ).
\end{equation}
The baryon number of this state is given by $B=(f(0)-f(\infty))/\pi=1$.
This solution breaks translational invariance, as well as the $O(4)$
spin-isospin symmetry. If we perform a global SU(2) isorotation on the
state,
\begin{equation}
U_1(\vec{r}|A) = A^\dagger U_1(\vec{r}) A,
\end{equation}
we obtain a state of the same energy. One can easily show that an
isorotation
$A$ has the same effect on $U$ as performing an ordinary rotation
 $R[A^{-1}]$
on $\hat{r}$, so that the grandspin $K$, the sum of spin and isospin, is a
good quantum number, with eigenvalue $0$,
\begin{equation}
\vec{K} = \vec{S} + \vec{I}.
\end{equation}

In the $B=2$ system we shall frequently use the product ansatz. This
ansatz makes use of the fact that the product of two $B=1$ solutions
has baryon number two. The most general ansatz we can therefore construct
from  two hedgehogs consists of the product of two rotated hedgehogs,
where the centers have been translated,
\begin{eqnarray}
U_2(r|\vec{R} AB)&=& A^\dagger U_1(\vec{r}-\vec{R}/2) A
                     B^\dagger U_1(\vec{r}+\vec{R}/2) B
\nonumber\\
                 &=& U_2(r|\vec{R} CD)
\nonumber\\
                 &=& D^\dagger C^{1/2\dagger} U_1(\vec{r}-\vec{R}/2) C
                     U_1(\vec{r}+\vec{R}/2) C^{1/2\dagger} D.
\label{eq:PA}
\end{eqnarray}
In the last of (\ref{eq:PA}) we have introduced the matrix $D$,
that describes the rigid isorotation of the whole system,
as well as a relative isorotation $C$. When $R$ is very large changing $C$
or
$D$ does not change the energy of the solution. For smaller $R$, $D$
still generates
a zero-mode (corresponding to broken isospin symmetry), but the energy will
depend on $C$. Again, the energy is also invariant under spatial rotation,
due to the conservation of angular momentum $\vec{J}=\vec{L}+\vec{S}$.

\subsection{Determination of manifold}
After having constructed all the constrained $B=2$
solutions, we construct a manifold that contains all these solutions.
This method, which is well know in nuclear structure physics
as the adiabatic large amplitude collective motion \cite{ALACM}, seems
first to have been introduced in the theory of interacting Skyrmions
by Manton \cite{MantonPRL}, under the name unstable manifold. His
discussion is  based on the deep geometric results
by Atiyah and Hitchin for the interaction of BPS monopoles
\cite{AtiyahHiggins}, which make
an identification of the structure of the manifold very clear but
do not give many tools for constructing it.

One of the ways to solve for the adiabatic collective manifold
would be to solve a local fluctuation equation. This is a very complex
problem for a field theory, but even when we reduce the complication by
selecting only a finite number of degrees of freedom as in the
Atiyah-Manton ansatz \cite{AM,HOA}
this remains a formidable (but interesting)
task.   The minimal set of degrees of freedom can be identified
without performing  any calculation, however.
There are six trivial ones,
consisting of the three center of mass coordinates, as well as global
rotations of the whole system (in both space and isospace), which we
shall ignore in the present discussion.
For large separations we find that an additional six degrees of freedom
are necessary to describe the product ansatz (\ref{eq:PA}):
the relative separation $\vec R$ (three coordinates)
and three more coordinates describing the relative orientation (as
discussed in the previous subsection usually represented by the $U(2)$
matrix $C = c_4I+i\vec{c} \cdot \vec{\tau}$, where $c_4^2 +\vec{c}^2=1$).
The minimal degrees of freedom are just the continuously deformed
set of six parameters as we take $R$ to smaller values. The meaning of the
word ``adiabatic'' in adiabatic manifold is that the motion in the
manifold
consists of the slowest modes. This
only makes sense if at any point all the
unstable directions lie within the collective space, so that no
 lower energy mode lies outside the manifold.
An analysis of the unstable $B=2$
hedgehog by Manton \cite{MantonPRL} shows that it is plausible that
this state has six unstable modes, again corresponding to relative
separation (three modes) and a relative rotation (another three). Since
this hedgehog is supposed to be part of the collective surface, we again
find a minimum of six collective modes.

Once one determines the collective surface there exist standard
techniques \cite{ALACM} to determine the collective Hamiltonian, with
the restriction that it is quadratic in momenta. Schematically it
takes the form
\begin{equation}
H = K(R,\hat{R},C,p_{R}, L_i, p_{Cj}) +
\frac{1}{2}I_iM_{ij}I_j + \frac{1}{2}J_iN_{ij}J_j
+ V(R, \hat R, C)
\end{equation}
Here $p_R$ is the momentum associated with $R$, $L$ the orbital
angular momentum and $p_C$ is shorthand for the momentum associated
with $C$. The symbols $I_i$ and $J_i$ denote the intrinsic components of
the (conserved)
isospin and angular momentum, respectively.
The kinetic energy is closely related to the geometric
structure of the collective surface, and can be very complicated. One
example that has been studied extensively is the case of the BPS
monopole, where one has only kinetic and no potential energy. It has
been proven \cite{AtiyahHiggins} that the underlying geometry
 is reasonably complicated and allows for  non-trivial
scattering of monopoles. In the current work we concentrate on
the potential energy, however. We assume that the kinetic terms
(including the contributions of $I$ and $J$) take the form
\begin{equation}
K = \frac{1}{2M} {\vec p}_R^2 + \frac{1}{2\cal I} (C_2^1 + C_2^2),
\end{equation}
where $C_2^i$ denotes the $O(4)$ Casimir operator $\mbox{$\frac{1}{2}$}
(S_i^2+I_i^2)$.
The only correction to this that has been studied is the $R$ and $C$
dependence of $M$ \cite{mass}, but we  ignore that in this paper.
It will be interesting to return to this problem in the context of the
$NN$ interaction.

To calculate the manifold beyond
the product ansatz we use the fact that for special choices of $C$,  due to
symmetry, there exist three stable situations \cite{HOA}
-- the stable attractive
channel, the marginally stable hedgehog-hedgehog and repulsive channels
-- where a path of the collective surface is traced out for fixed $C$.
In each of these channels (labeled by $i$ for convenience) one
calculates the classical potential energy $V_i$. The identification of
the symbols $R$ in each of the three Hamiltonians as one and the same
radial coordinate is a bit precarious. There is no apparent reason why
we cannot separately redefine the radial coordinate in each channel
(say, by a point transformation). Here the topology comes into play
however. Since each channel corresponds to a definite choice of $C$,
the three point
transformations can be embedded in a single transformation that depends
on both $R$ and $C$,
so that we are just making a change of coordinate
transformation on the manifold, by which we do not gain much. In order
to sensibly identify the $R$'s in the same channel we should use
geometric notions. First note that each of the
three channels $i$ seems to correspond to a geodesic line on the
collective surface, due to the extra symmetry of the pion field. In
order to be able to make a meaningful identification of points on the
three lines as having the same $R$ the three points
should lie on a single geodesic surface on the manifold, that crosses
the three geodesic lines $i$ orthogonally. This is of course extremely
difficult to implement, and we do not know of a good way of doing this.
In most discussions of the Skyrmion-Skyrmion interaction, including the
present one, these problems
are ignored, and a special definition of $R$ is made that is then used
to identify the different channels.

\subsection{The algebraic model}
Having determined the functions $V_i$ we  need to understand their
relation to the nucleon-nucleon potential. The usual approach is to
return to the product ansatz, where one makes very strong assumptions
about the form of the two-Skyrmion solution. Again, this form is exact
in the large-$R$ limit. The advantage of this form is that it guides
us into a particular expansion of $V$ in terms of functions
$\vec{c}\cdot\hat{R}$, $c_4$ (the quaternion $c$ specifies the relative
orientation). This expansion is closely related to an algebraic model
of the Skyrmion \cite{OBBA}, which in the large-$N$ limit behaves in
exactly this manner. There is no a priori reason why this form should
be restricted to the product ansatz, however. As we have argued above
the coordinates of the collective surface are in one-to-one
correspondence with those of the product ansatz, so that the current
problem can again be modeled by a similar form \cite{remarkdoughnut}.

Let us now discuss the algebraic model \cite{ABO,OBBA} that we use.
Originally it was introduced to give an algebraic alternative
to the rotational zero-mode quantization of the Skyrmion. Since the
Skyrmion breaks   $SO(4) \sim SU_I(2) \times SU_S(2)$ symmetry, but is
invariant under a $SO(3)$ subgroup, we find that we can describe each of
the equivalent configurations by a point of on the sphere $S_3
\sim SO(4)/SO(3)$ \cite{remarkNP}.

We can now introduce a bosonic model with dynamical symmetry $SU(4)$,
which in the limit of large boson number is also described by a point
on the sphere $S_3$. Without discussing the hamiltonian in any detail,
we can draw some interesting conclusions. For any $N$ (the number of
bosons) we find that, as in the Skyrme model, all states have $I=S$.
The  spectrum of the algebraic model is cut off at $I=N/2$, however.
Since we believe that in the real world, where
$N_{\rm C}=3$, we have a cut-off at isospin $3/2$ for the delta, we
impose
the condition
that the number of bosons $N$ is equal to the number of colors
$N_{\rm C}$. If
we make this identification one can immediately show that
for $N=N_{\rm C}=3$ the algebraic model is
identical to the non-relativistic quark model without radial
excitations. It has been shown that this model can be used to extract
finite $N_{\rm C}$ corrections to observables. Of course we quantize
only a few coordinates and  we can only calculate a very limited set of
finite $N_{\rm C}$ corrections, corresponding to the zero-mode
quantization.
One may hope, however, that through a judicious choice of collective
coordinates we have found the most important, and maybe even the
largest, contributions.

In this paper
we are not interested in the single Skyrmion case, but rather in the
the case of two Skyrmions. This has been discussed in quite some detail
for the case of the product ansatz in Ref.\ \cite{OBBA}. The algebraic
model used consists of two sets of $u(4)$ algebras, one for each
Skyrmion, as well as a radial coordinate $\vec{R}$. The interaction
of the two Skyrmions can be expanded in terms of three operators,
the identity $I$, and the operators $W$ and $Z$:
\begin{eqnarray}
W & = & T^{\alpha}_{pi} T^{\beta}_{pi}/N_{\rm C}^2,\nonumber\\
Z & = & T^{\alpha}_{pi} T^{\beta}_{pj}
[3\hat{R}_i \hat{R}_j-\delta_{ij}]/N_{\rm C}^2.
\end{eqnarray}
Here $\alpha$ and $\beta$ label
two different sets of bosons, used to realize the $u(4)$ algebras,
and $T$ is a one-body operator with
spin and isospin $1$. The semi-classical (large-$N_{\rm C}$) limit of these
operators can be given in terms of $\hat{R}$ and $C=c_4I+
\vec{c}\cdot\vec{\tau}$ as \cite{OBBA}:
\begin{eqnarray}
W_{\rm cl} & = & 3c_4^2-\vec{c}^2,\nonumber\\
Z_{\rm cl} &=& 6(\vec{c}\cdot\hat{R})^2-2\vec{c}^2    .
\end{eqnarray}
  If we expand the Skyrmion-Skyrmion interaction  we  find
\begin{equation}
v(R) = v_1(R) + v_2(R) W_{\rm cl} + v_3(R)Z_{\rm cl}+ {\rm higher\ order}.
\label{eq:pots}
\end{equation}
Fortunately studies of the $B=2$ system with the product ansatz
\cite{Jackson,VinhMau,OBBA},
the Atiyah-Manton ansatz  \cite{AM,HOA} and with
direct numerical integration of the equations \cite{WW} have shown that
much information about the $B=2$ energy as a function of separation
$R$ is
contained in the particular choice of three paths discussed earlier,
corresponding
to   three definite choices of the relative orientation, or
equivalently to   a special symmetry of the pion field.
 This allows us
to calculate an approximate form for the interaction energy of two
Skyrmions which consist of the first three term of (\ref{eq:pots})
 (the form used in Ref.~\cite{WW} is fully equivalent),
which correspond to an expansion up to first order only in terms of the
two scalar operators $W$ and $Z$.
One of the important consequences of the
explicit realization of these operators is that now
algebraic methods \cite{OBBA,ABO} can be used to obtain their
                         {\em explicit} finite $N_{\rm
C}$ realization. In this way we can include some (arguably the most
important) finite $N_{\rm C}$ corrections in the calculation.
This includes the well-known amplification of the
nucleon-nucleon matrix element of $T$
by the  factor $(N_{\rm C}+2)/N_{\rm C}$.

\section{The adiabatic interaction}

The problem
we address here is how to obtain a nucleon-nucleon interaction which
includes some finite $N_{\rm C}$ effects, from (\ref{eq:pots}).  To date
most workers have simply sandwiched (\ref{eq:pots})
between states containing two
nucleons for each value of $R$ and called that the interaction.
This is definitely not the nucleon-nucleon interaction determined
phenomenologically from scattering phaseshifts, where we can only
require that the state contains two nucleons at large separation.  For
shorter distances, when the hadrons are interacting they can be
(virtually) what ever the dynamics requires, for example $\Delta$'s. We
know \cite{HOA} that the baryon density is strongly
deformed for moderate $R$ -- a deformation that finally results in the
toroidal shape with its $70$ MeV of binding.  That configuration
corresponds to a great deal of mixing between all states of the
individual Skyrmions. This admixture will lead to additional attraction
at intermediate distances.  The purpose of this paper is to calculate
the effect of this mixing on the N-N
interaction for $N_{\rm C}=3$ starting with
(\ref{eq:pots}).
The need for state mixing has been realized before \cite{Delta} in the
context of the product ansatz, but that ansatz is a poor starting point.
Furthermore, as we state below,  state mixing only makes sense for
finite $N_{\rm C}$. At the same time
there are algebraic finite $N_{\rm C}$  corrections
to matrix elements that are difficult to include in the formalism of
\cite{Delta}. For these reasons, these early attempts to introduce state
mixing did not find sufficient central attraction.
We should remark here that a very different formalism, based on a study
of fluctuations around the product ansatz \cite{Zahed} has been used
to show that some attraction exists in two-pion range. Unfortunately
the expansion underlying this approach fails in the region that
is most crucial to central attraction; 1 fm $<R<$ 2 fm.

For large $N_{\rm C}$, one can distinguish two
energy scales or reciprocally two time scales. The  slower time scale is
associated with the motion in the collective manifold, i.e., $R$ and the
orientation. The other time scale corresponds to the almost
instantaneous response of the
pion field to changes in $R$
and the relative isospin orientation.
For large $N_{\rm C}$ we cannot separate the time scales
for the two sets
of adiabatic modes, as can be seen in the highly correlated doughnut.
For $N_{\rm C}$ equal to three the situation changes. The $R$ motion is
typically much slower than the rotational motion which leads to the
separation of the nucleon and $\Delta$ states. We thus have three energy
scales,
that of the pion field, of the $N-\Delta$ separation and of the $R$ motion.
We can now calculate a Born-Oppenheimer potential for the
$R$-motion, which constitutes the slowest degree of freedom.

\subsection{Perturbative approach}
All of the effects
of quick response of the pion field at each $R$ are already in Eq.\
(\ref{eq:pots}).  It is the effect of the rotational states we want to
include.  As a guide, we begin by studying it perturbatively. Let us
call the full potential between two nucleons, including the effects of
the rotational excited states, $V$. In terms of the Skyrmion interaction
$v$ of (\ref{eq:pots}) and to second order, $V$ is given by
\begin{equation}
V(R)=\langle NN|v(R)|NN\rangle  + \sum_s~\!\!'\; \frac{\langle
NN|v(R)|s\rangle  \langle s|v(R)|NN\rangle }{E_{NN}(R) -E_{s}(R)} .
\label{eq:no4}
\end{equation}
(Here $E_{NN}$ is the two-nucleon energy and $E_S$ is the energy of
the relevant excited state.)
The first term on the right is the direct nucleon-nucleon projection of
$v$ and is the term that has appeared in the literature.  The second
term is  the correction due to rotational or excited states.
It is clear from
the energy denominator that the second term is attractive.
For
$N_{\rm C}=3$ the states $|s\rangle $ that can
enter are $|N \Delta\rangle ,
|\Delta N \rangle $ and   $|\Delta \Delta\rangle $.
  Except   for a centrifugal term we shall add later,
the excitation energy is assumed to be independent of $R$,
and thus can be
expressed in terms of the N-$\Delta$ energy difference, $300$ MeV.
It is here that we are neglecting the effect of the position dependence of
the inertial parameters.

The lowest order potential has been calculated by several groups
using several levels of approximation to the Skyrme model
\cite{VinhMau,WW,HOA}.
The form found (or assumed), however, can always be cast in the
form of the algebraic model given by \cite{OBBA}. This allows us to
study both the large-$N_{\rm C}$ limit, as well as to include finite
$N_{\rm C}$ effects
explicitly in a systematic way. (The reader should note that the results
in \cite{WW} include corrections for the tensor  and spin-spin
interactions.)

The leading term in this expansion is given by the form
(cf.~(\ref{eq:pots}))
\begin{equation}
v(\vec R) = {v_1(r)}
 + {v_2(R)}W + {v_3(R)} Z(\hat R).
\label{eq:2p1}
\end{equation}
The algebraic operators $W$ and $Z$
have very simple expectation value for nucleons,
(see \cite{OBBA})
\begin{eqnarray}
\left<NN\right| W \left|NN\right>& = & \mbox{$\frac{1}{9}$} P_N^2
\left<NN\right|
\sigma^1\!\cdot \sigma^2 \tau^1\!\cdot\tau^2
\left|NN\right>,\nonumber\\
\left<NN\right| Z \left|NN\right>& = & \mbox{$\frac{1}{9}$}  P_N^2
\left<NN\right|     (3\sigma^1\!\cdot \hat R\;\sigma^2 \!\cdot\hat R-
              \sigma^1\!\cdot \sigma^2 )\tau^1\!\cdot\tau^2
\left|NN\right> ,
\end{eqnarray}
{\em i.e.}, apart from a finite $N_{\rm C}$ correction factor $P_N^2$
($P_N = 1 +2/N_{\rm C}$) they represent the
spin-spin and tensor interactions.
If we use this form we find that the lowest order interaction in
the nucleon-nucleon channel is given by
\begin{equation}
V^{(0)} = v_1 + \frac{v_2P_N^2}{9}\sigma^1\!\cdot \sigma^2
\tau^1\!\cdot\tau^2
              + \frac{v_3P_N^2}{9}
             (3\sigma^1\!\cdot \hat R\sigma^2 \!\cdot\hat R-
              \sigma^1\!\cdot \sigma^2 )\tau^1\!\cdot\tau^2.
\label{eq:V0}
\end{equation}

If we wish to evaluate the perturbation correction (\ref{eq:no4}) we must
evaluate the two sets of matrix elements
$\left<NN\right| \tilde v(\vec R) \left|N\Delta\right>
\left<N\Delta\right| \tilde v(\vec R) \left|NN\right> $ and
$\left<NN\right| \tilde v(\vec R) \left|\Delta\Delta\right>
\left<\Delta\Delta\right| \tilde v(\vec R) \left|NN\right>$
separately. Details are discussed in appendix A, but the final
result for the first order correction to the  NN interaction is
\begin{eqnarray}
V^{(1)}_{\rm PT} & = &  -\frac{Q_N^2}{\delta}\{
[\mbox{$\frac{1}{3}$}Q_N^2 P^\tau_0 + (\mbox{$\frac{16}{27}$}P_N^2+
\mbox{$\frac{5}{27}$}Q_N^2)
P^\tau_1]
(v_2^2+2v_3^2) + \nonumber\\
&&   (\sigma^1\!\cdot \sigma^2 )
[-\mbox{$\frac{1}{18}$}Q_N^2 P^\tau_0+
(\mbox{$\frac{16}{81}$}P_N^2-\mbox{$\frac{5}{162}$}Q_N^2)P^\tau_1]
(v_2^2-v_3^2)\nonumber\\
&&
(3\sigma^1\!\cdot \hat R\sigma^2 \!\cdot\hat R-\sigma^1\!\cdot\sigma^2 )
[-\mbox{$\frac{1}{18}$}Q_N^2 P^\tau_0+
(\mbox{$\frac{16}{81}$}P_N^2-\mbox{$\frac{5}{162}$}Q_N^2)P^\tau_1]
(v_3^2-v_2v_3)\}.
\label{eq:PT}
\end{eqnarray}
Here $Q_N$ is an additional finite $N_{\rm C}$ correction factor
\begin{equation}
Q_N = \sqrt{(1-1/N_{\rm C})(1+5/N_{\rm C})},
\end{equation}
$\delta$ is the N-$\Delta$ energy difference, and $P^\tau_T$ is a
projection operator onto isospin $T$.

\subsection{Diagonalization of the interaction}
The perturbative expansion discussed in the previous subsection just
gives a rough estimate of the effect of state mixing in  the
Born-Oppenheimer
approximation. To solve for the Born-Oppenheimer Hamiltonian
the total collective Hamiltonian is first divided into two parts,
an intrinsic part and a part that only describes the radial motion,
\begin{equation}
H = -\frac{\hbar^2}{2M}R^2\partial_R R^{-2}\partial_R + \tilde{H}(R).
\end{equation}
The hamiltonian $\tilde{H}(R)$ is nothing but the sum of the potential
$v(R)$ we used in the perturbative calculation, plus the kinetic term
used to calculate the energy difference in the energy denominators,
\begin{equation}
K = \frac{1}{2\Lambda}(I_1^2+I_2^2) + \frac{\hbar^2}{2MR^2} L(L+1).
\end{equation}

We now  introduce the instantaneous eigenstates of the intrinsic
Hamiltonian.  For fixed $R$ we diagonalize $\tilde{H}$,
\begin{equation}
\tilde{H}(R)\left|i\right>_R = E_i(R) \left|i\right>_R.
\end{equation}
The standard Born-Oppenheimer
 approach is  to start  at large $R$, where the states
are pure $NN$, and follow the continuous energy curve as $R$ becomes
smaller to define the adiabatic potential energy. This is basically
the procedure we follow, apart from a few additional complications
encountered if we have more than one $NN$ state in any given channel.
This is solved by using a model Hamiltonian to determine the unmixed
adiabatic eigenstates, as will be discussed below.

Having defined the adiabatic eigenstates, the mathematically correct
way to derive the adiabatic Hamiltonian is to decompose a general
time-dependent solution of the Schr\"odinger equation in adiabatic
eigenstates,
\begin{equation}
\left|\Psi(t)\right> = \sum_i a_i(R,t) \left|i\right>_R.
\end{equation}
If we  substitute this in the time-dependent Schr\"odinger equation,
and multiply from the left with $_R\!\left<i\right|$, we find
\begin{eqnarray}
i \dot{a}_i \delta_{ij}& =  &\delta_{ij} \left(\frac{-\hbar^2}{2M}
\frac{1}{R^2}\partial_R R^2\partial_R
 +E_i(R)\right) a_i
\nonumber\\&&
+\frac{-\hbar^2}{M}
 \sum_j~ _R\left<i\right| \partial_R \left|j\right>_R \partial_R a_j(R,t)
\nonumber \\&&
+ \sum_j~ _R\left<i\right| \frac{-\hbar^2}{2M}
\frac{1}{R^2}\partial_R R^2\partial_R \left|j\right>_R a_j(R,t).
\end{eqnarray}
This expression exhibits both the adiabatic Hamiltonian (the first
term on the right-hand side), and the corrections. If we assume there
are no diagonal terms of $\partial_R$ between adiabatic eigenstates
(since there are no closed circuits in $R$ space, it seems a reasonable
assumption to ignore Berry's phase), the first correction describes a
velocity dependent transition potential. The next term gives a non-adiabatic
correction to the diagonal adiabatic  potential, as well as a velocity
independent transition potential.

\subsubsection{The potential in angular momentum coupled form}
     To evaluate (\ref{eq:no4}) it is convenient to introduce states
$|I_1I_2LSJT\rangle$ labeled by the conserved (total) isospin $T$ and
angular momentum $J$, as well as the  NN channel orbital angular momentum
$L$ and spin $S$.
Since we wish to evaluate the matrix elements of $\tilde{H}(R)$
 in the basis $\left|I'_1 I'_2 LSJT\right>$ we need to know the
matrix elements of the potential $v$ in this basis.
After performing some standard
angular momentum algebra, we find that
\begin{eqnarray}
\lefteqn{\left<I_1I_2LSJT\right| v   \left|I'_1 I'_2 LSJT\right>}
\nonumber\\
& = &
v_1 \delta_{SS'} \delta_{LL'} \delta_{I_1I'_1} \delta_{I_2I'_2}
\nonumber\\
& & +
\frac{v_2}{9}
(-1)^{S+T}
     \delta_{SS'} \delta_{LL'}
 {\setlength{\arraycolsep}{2pt}\left\{
\begin{array}{ccc}
{\scriptstyle I_1}&{\scriptstyle I_2}&{\scriptstyle S}\\
{\scriptstyle I'_2}&{\scriptstyle I'_1}&{\scriptstyle 1}
\end{array}\right\}}
 {\setlength{\arraycolsep}{2pt}
\left\{\begin{array}{ccc}
{\scriptstyle I_1}&{\scriptstyle I_2}&{\scriptstyle T}\\
{\scriptstyle I'_2}&{\scriptstyle I'_1}&{\scriptstyle 1}
\end{array}\right\}}
\left.\left.\left<I_1\right|\right|\right|T^{(11)}
\left.\left.\left|I'_1\right>\right.\right.
\left.\left.\left<I_2\right|\right|\right| T^{(11)}
\left.\left.\left|I'_2\right>\right.\right.
\nonumber\\
& & +
 \frac{      {v}_3}{9}
  \sqrt{30}
(-1)^{L+L'+S+J+T+I_2+I_1'}
\hat{L} \hat{L}' \hat S \hat{S}'
\nonumber \\
&&    \times
 {\setlength{\arraycolsep}{2pt}
\left(\begin{array}{ccc}
{\scriptstyle L}&{\scriptstyle 2}&{\scriptstyle L'}\\
{\scriptstyle 0}&{\scriptstyle 0}&{\scriptstyle 0}\end{array}\right)}
 {\setlength{\arraycolsep}{2pt}
\left\{\begin{array}{ccc}
{\scriptstyle S}&{\scriptstyle L}&{\scriptstyle J}\\
{\scriptstyle L'}&{\scriptstyle S'}&{\scriptstyle 2}
\end{array}\right\}}
 {\setlength{\arraycolsep}{2pt}
\left\{\begin{array}{ccc}
{\scriptstyle I_1}&{\scriptstyle I_2}&{\scriptstyle T}
\\{\scriptstyle I'_2}&{\scriptstyle I'_1}&{\scriptstyle 1}
\end{array}\right\}}
 {\setlength{\arraycolsep}{2pt}\left\{\begin{array}{ccc}
{\scriptstyle I_1}&{\scriptstyle I_2}&{\scriptstyle S}\\
{\scriptstyle I'_1}&{\scriptstyle I'_2}&{\scriptstyle S'}\\
{\scriptstyle 1}&{\scriptstyle 1}&{\scriptstyle 2}\end{array}\right\}}
\left.\left.\left<I_1\right|\right|\right| T^{(11)}
\left.\left.\left|I'_1\right>\right.\right.
\left.\left.\left<I_2\right|\right|\right|T^{(11)}
\left.\left.\left|I'_2\right>\right.\right..
\end{eqnarray}
The relevant reduced matrix elements of $T$ are
\begin{eqnarray}
\left.\left.\left<N\right|\right|\right| T^{(11)}
\left.\left.\left|N\right>\right.\right.
&=& -10, \\
\left.\left.\left<\Delta\right|\right|\right| T^{(11)}
 \left.\left.\left|\Delta\right>\right.\right.
&=& -20, \\
\left.\left.\left<\Delta\right|\right|\right| T^{(11)}
 \left.\left.\left|N\right>\right.\right.
&=& -8\sqrt{2}.
\end{eqnarray}
The matrix elements of the kinetic part are taken to be very simple,
\begin{equation}
\left<I_1I_2LSJT\right|K  \left|I'_1 I'_2 L'S'JT\right> =
\delta_{I_1I'_1}\delta_{I_2I'_2}\delta_{LL'}\delta_{SS'}
\left(300 [I_1+I_2-1/2] +\frac{L(L+1)}{2M_{I_1I_2}R^2}\right).
\end{equation}
 To account for the effects of centripetal
repulsion at small $R$, we have added the
centrifugal energy, $      {\hbar^2}         L(L+1)/ 2M_{I_1I_2}R^2$,
to channel energies.  We take $M_{I_1I_2}$ to be the reduced
mass in the relevant channel,
\begin{equation}
M_{I_1I_2} = \frac{M_{I_1}M_{I_2}}{M_{I_1}+M_{I_2}},
\end{equation}
where $M_{1/2} = 932$~MeV and $M_{3/2} = 1232$~MeV.

For sake of comparison we shall also need the matrix elements of a
model nucleon-nucleon interaction in the same channel (we have suppressed
the spin and isospin projection quantum numbers),
\begin{eqnarray}
\lefteqn{%
\left<NNLSJT\right|%
V_{\rm c}^T + V_{\rm s}^T \sigma^1 \cdot \sigma^2%
+ V_{\rm t}^T \sigma^1_i \sigma^2_j (3\hat{R}_i\hat{R}_j%
                 - \delta_{ij})%
\left|NNL'S'JT\right>}  \nonumber\\
& = & \delta_{SS'} \delta_{LL'}
\left(V_{\rm c}^T+
V_{\rm s}^T (-1)^{1+S} 6
{\setlength{\arraycolsep}{2pt}\left\{\begin{array}{ccc}
{\scriptstyle 1/2}&{\scriptstyle 1/2}&{\scriptstyle S}\\
{\scriptstyle 1/2}&{\scriptstyle 1/2}&{\scriptstyle 1}
\end{array}\right\}} \right) +
\nonumber\\&&
  6 \sqrt{30}         V_{\rm t}^T
(-1)^{L+L'+S+J} \hat L \hat{L'} \hat{S}\hat{S'}
 {\setlength{\arraycolsep}{2pt}\left(\begin{array}{ccc}
{\scriptstyle L}&{\scriptstyle 2}&{\scriptstyle L'}\\
{\scriptstyle 0}&{\scriptstyle 0}&{\scriptstyle 0}\end{array}\right)}
 {\setlength{\arraycolsep}{2pt}\left\{\begin{array}{ccc}
{\scriptstyle L}&{\scriptstyle S}&{\scriptstyle J}\\
{\scriptstyle S'}&{\scriptstyle L'}&{\scriptstyle 2}\end{array}\right\}}
 {\setlength{\arraycolsep}{2pt}\left\{\begin{array}{ccc}
{\scriptstyle 1/2}&{\scriptstyle 1/2}&{\scriptstyle S}\\
{\scriptstyle 1/2}&{\scriptstyle 1/2}&{\scriptstyle S'}\\
{\scriptstyle 1}&{\scriptstyle 1}&{\scriptstyle 2}\end{array}\right\}}.
\label{eq:Vfit}
\end{eqnarray}
Here we have chosen to use a parametrization identical to the one used
in the discussion of perturbation theory.

In order to determine an effective potential from the analysis given
above we perform a calculation that is standard in the
Born-Oppenheimer approximation. We  assume that
at each $R$ we can diagonalize the potential. We then follow a state
from large $R$, where it is purely made up out of  nucleons, to small
$R$ where we have mixing. This gives the usual adiabatic potential curves
of molecular physics. A complication will be that we may have more than
one nucleon state for given quantum numbers. We shall see that we can
still fit the interaction in that case.

The model space splits into two parts that are either symmetric or
antisymmetric under interchange of the particles \cite{Note}. We select
the antisymmetric part where the NN states satisfy the standard selection
rule $L+S+T$ is odd. Of course, in contrast to the lowest order
perturbative result, we
will now find a channel dependent potential.

The fact that we have a different reduced mass in each channel will
also lead to a position dependent effective mass. Suppose that
$c_{I_1I_2}^2(R)$ gives the percentage of the ${I_1I_2}$ state in
the diagonalization of the Born-Oppenheimer Hamiltonian.
In that case the average radial kinetic energy is
\begin{equation}
\sum_{I_1I2} -\frac{\hbar^2}{2M_{I_1I_2}}  c_{I_1I_2}^2(R)
\frac{1}{R^2} \partial_R R^2 \partial_R,
\end{equation}
and we can define a position dependent mass by
\begin{equation}
M(R) = 1/\left(\sum_{I_1I2} \frac{c_{I_1I_2}^2(R)}{M_{I_1I_2}}\right).
\label{eq:MR}
\end{equation}

\subsubsection{Derivation of the potential}

Let us first consider the case $T=0$. A little analysis shows that
in this case the two channels  $J^\pi=1^+$ and $J^\pi=1^-$ give
three two-nucleon states, enough to determine the three independent
functions in (\ref{eq:Vfit}). (There are no $NN$ states with
$T=0$ and $J=0$.)

For $J^\pi=1^+$ we have the two possible states
$\left|NNL=0S=1J=1T=0\right>$ and $\left|NNL=2S=1J=1T=0\right>$.
The matrix elements
of (\ref{eq:Vfit}) in this space take the form
\begin{equation}
\left( \begin{array}{cc}
V_{\rm c}^0+V_{\rm s}^0   & 2\sqrt{2} V_{\rm t} \\
2\sqrt{2} V_{\rm t} &  V_{\rm c}^0+V_{\rm s}^0 -2V_{\rm t}^0
\end{array} \right),
\end{equation}
with eigenvalues
\begin{equation}
E_1 = V_{\rm c}^0+V_{\rm s}^0 +2V_{\rm t}^0,\;\;\;
E_2 = V_{\rm c}^0+V_{\rm s}^0 -4V_{\rm t}^0.
\end{equation}
These numbers should be fitted to the two lowest eigenvalues of the
diagonalization of the algebraic potential in the complete nucleon-delta
space. If we also look at the state
$\left|NNL=1S=0J=1T=0\right>$, with matrix element of
 (\ref{eq:Vfit})  equal to
\begin{equation}
E_3 = V_{\rm c}^0  -3V_{\rm s}^0,
\end{equation}
(again, this is equated to the lowest eigenvalue of the complete
diagonalization) we find
\begin{eqnarray}
        V_{\rm c}^0 & = &  E_1/2+E_2/4+E_3/4, \\
        V_{\rm s}^0 & = & E_1/6+E_2/12-E_3/4, \\
        V_{\rm t}^0 & = & (E_3-E_2)/6.
\end{eqnarray}

A similar calculation for $T=1$ involves the three channels
$J^\pi=0^+,0^-,1^-$ (again $L+S+T=$odd).
The relevant matrix elements of (\ref{eq:Vfit}) are
\begin{eqnarray}
E_1'  & \equiv &
 \left<NNL=0S=0J=0T=1\right|  V
 \left|NNL=0S=0J=0T=1\right>
 \nonumber \\ & = &
    V_{\rm c}^1-3V_{\rm s}^1 \nonumber\\
E_2'  & \equiv &
 \left<NNL=1S=1J=1T=1\right|  V
 \left|NNL=1S=1J=1T=1\right>
 \nonumber \\ & = &
  V_{\rm c}^1+V_{\rm s}^1+2V_{\rm t}^1 \nonumber\\
E_3'  & \equiv &
 \left<NNL=1S=1J=0T=1\right|  V
 \left|NNL=1S=1J=0T=1\right>
 \nonumber \\ & = &
  V_{\rm c}^1+V_{\rm s}^1-4V_{\rm t}^1 .
\end{eqnarray}
Thus
\begin{eqnarray}
V_{\rm c}^1&=&E'_1/4+E'_2/2+E'_3/4, \nonumber \\
V_{\rm s}^1&=&-E'_1/4+E'_2/6+E'_3/12, \nonumber \\
V_{\rm t}^1&=&       (E'_2-E'_3)/6.
\end{eqnarray}

We  subtract the  centrifugal force from our potentials,
since in general this is not taken to be included in the potential itself.
We had to include it in our kinetic term, since we make the Born-Oppenheimer
separation between the radial motion and everything else, including the
orbital motion.

\section{Results}
As stated before,
   for each total isospin, $T=0,1$ we write $V$ in the $|NN LSJT\rangle
$ space as
\begin{equation}
\langle NN LSJT| V_{\rm c}^{T} + V_{\rm s}^{T} \sigma^{1} \cdot
\sigma^{2} +
  V_{\rm t}^{T} \sigma_{i}^{1} \sigma_{j}^{2} (3 \hat{R}_{i} \hat{R}_{j}
     - \delta_{ij}) |NN L'SJT\rangle .
\end{equation}
It is the $V_{\rm c}^{T}$, $V_{\rm s}^{T}$ and $V_{\rm t}^{T}$ that we
study.  We have extracted the necessary values of $v_1,v_2,v_3$ from the
work of Walhout and Wambach \cite{WW}. This is an uncertain process
for large $R$ since the potentials are quite small in this region and
are calculated in \cite{WW} on a lattice by first calculating the
total interaction energy of the $B=2$ system and then subtracting
the rest energy of two free Skyrmions.
We wish to compare the Skyrme based potential with a realistic
nucleon-nucleon interaction.
One cannot relate our result to modern effective potentials, such as
the Bonn \cite{Bonn} and Paris \cite{Paris} potentials, since their
central parts contain explicit momentum dependent terms. For that reason
we compare our potentials to the Reid soft core (RSC) potential
\cite{Reid}.

In
Figs.~1 and 2 we show the central potentials $V_{\rm c}^T$ calculated from
Eq.\ (\ref{eq:no4}), first from the first
term on the right of (\ref{eq:no4}) only (this is the result of Ref.\
\cite{WW}) and then with the Born-Oppenheimer method. Note we only show
the result for intermediate $R$, $1\ \mbox{fm}<R<2\ \mbox{fm}$.
To keep the figures simple we separate $T=0$ and $T=1$. The nucleons only
result from \cite{WW} is independent of $T$ and shows a weak attraction.
The perturbation result (\ref{eq:PT}) gives considerably more
attraction, which is in turn somewhat reduced by the full Born-Oppenheimer
(BO) diagonalization. This reduction comes mostly from the centrifugal
terms in the kinetic energy which were not included in the perturbative
calculations, and partly from the higher order effects included in the
diagonalization.
The centrifugal terms reduce the mixing at small
$R$, and hence lead to less central attraction.
All the results, nucleons only, perturbation theory and BO diagonalization
agree at large $R$. This reflects the fact that state mixing involves
multiple pion exchange and hence has intermediate range. Also
shown in Figs.~1 and 2 are the $T=0$ and $T=1$ components of the
Reid soft core potential. They have less central attraction than we
find. As we shall see that reflects a rather different shuffling of
attraction between the central and the spin dependent interactions
between the RSC and the Skyrme approach. But even at this level
the fact that the combination of careful solution of the $B=2$ sector,
state mixing and finite $N_{\rm C}$ effects gives {\it too much}
attraction is a welcome change from the many previous attempts to
derive the $NN$ interaction from the Skyrme approach which always
found too little central attraction.

In Figs.~3 and 4 we show the $T=0$ and $T=1$ spin-dependent potentials.
In these cases the nucleons-only potential, the perturbative results
and the full diagonalization are all quite similar, but
now the RSC is stronger than the Skyrme result, rather than weaker as it
was in the central potential. This is the shuffling we referred to above.
Figs.~5 and 6 show the tensor force for both values of isospin.
Except for $T=1$ at small $R$ all the Skyrmion calculations
and even the RSC agree. This reflects the dominance of one
pion exchange for the tensor potential.

 We see  that though
the effect of the Born-Oppenheimer corrections are modest for large $R$ they
become large for intermediate $R$ in the central channel,
but remain small in the other channels. The reason for this is that the
corrections are small for all channels compared to the typical size of the
potential, but that the lowest order central potential is much smaller
than this typical size.

We saw in Figs.~1--4 that the RSC gives less central spin-independent
attraction than we find but a stronger spin-dependent potential.
It is therefore instructive to compare the more physical $S$-channel
$L=0$ potentials. In Fig.~7 and 8 we show the $^3S_1$ ($T=0$) and
the $^1S_0$ ($T=1$) potentials. Now we see that the full Born-Oppenheimer
potential looks quite similar to the RSC while the one obtained
without mixing does not resemble the RSC.
It is in this sense that we say that the present results give a credible
account of the nucleon-nucleon
interaction at intermediate range.

We now turn to a more detailed analysis of the Born-Oppenheimer
approximation, the adiabatic approximation and non-adiabatic corrections.
We analyze a representative case, the $J=1^-$, $T=1$ channel, since
it mixes $NN$, $N\Delta$ and  $\Delta\Delta$ states.
The relevant states in the diagonalization
are $\left|NN,L=1,S=1\right>$, $\left|N\Delta,L=1,S=1\right>$,
$\left|N\Delta,L=1,S=2\right>$,
$\left|\Delta\Delta,L=1,S=1\right>$ and
$\left|\Delta\Delta,L=1,S=3\right>$. In Fig.~9 we plot
the energy of the adiabatic eigenstates, which remain distinct for
all separations, except for a narrow avoided crossing near
$1.2$~fm.
For this channel all the energies are repulsive. In Fig.~10 we show the
percentage of each component in the lowest adiabatic eigenstate.
It is this state that was used in the determination of the $NN$ potential.
 One
can see that the $NN$ component is dominant, and we have only small
admixtures
of the other components. For small $R$ it is the third adiabatic
eigenstate that admixes most strongly.

In Fig.~11 we show the effective mass of the lowest adiabatic eigenstate
as defined by Eq.~(\ref{eq:MR}). Since the only effect on the
masses we include is the centrifugal term, the mass increases with
decreasing $R$. Preliminary calculations of the mass parameters for the
classical $B=2$ system give mass parameters that decrease with
decreasing $R$, at least in the attractive channel \cite{mass}. Thus
the results of Fig.~11 should be taken only to
remind us that the Skyrme approach leads to
a picture of nucleons that distort and change their mass as they
interact.

Corrections to the adiabatic approximation lead to transition potentials
among the adiabats. The relevant matrix elements
 can be calculated exactly using standard
algebraic techniques:
\begin{equation}
_R\left<i\right|\partial_R \left|j\right>_R
=~_R\left<i\right|(\partial_R\tilde H(R)) \left|j\right>_R/
(E_i(R)-E_j(R)).
\end{equation}
We further assume that the (imaginary) diagonal matrix element is zero, which
corresponds to a consistent real choice of $\left|j\right>_R$.
The second derivatives can now be evaluated using a sum over intermediate
states. As argued in the previous section we have two terms, a local
 transition potential and a velocity dependent potential.
We continue to study the J=$1^-$, $T=1$ channel
and consider the transition from the $NN$ adiabat to each of the five
others. The local potentials for these transitions are shown in Fig.~12
as a function of $R$.
They are on average much smaller than the difference between the
relevant adiabatic eigenvalues, as they should be if the adiabatic
approximation works.
Near $R=1.2~{\rm fm}$ there is a spectacular excursion of the velocity
dependent potential in Fig.~12. This can be traced back to the
narrowly avoided level crossing in Fig.~9.
This picture would suggest that at high $NN$ energy there should be the
onset of strong
$\Delta$ production. The theory is still too crude to make
a detailed prediction, but this aspect of $NN$ dynamics suggested
by the Skyrme approach deserves further investigation.
 The velocity dependent potential are a little
more difficult to interpret. We have chosen to plot, in Fig.~13, the
potential such that we have to multiply it by $-i$ times the scaled
radial velocity
$-i\beta=v/c = \frac{\hbar}{Mc} \partial_R$. As can be seen this potential
is very small for non-relativistic velocities.

\section{Summary and conclusions}

Here we have shown that the Skyrme model can give
strong mid-range nucleon-nucleon attraction that is in qualitative
agreement with
phenomenological potentials.  This is in sharp contrast
to early results based on the
product ansatz.
Two components play a key role in obtaining the attraction. First
one must pay attention to
the non-linear nature of the Skyrme Lagrangian, as
shown in Refs.\ \cite{WW,HOA}, but second one must also include configuration
mixing at intermediate distances as we have stressed here.  This mixing
is not easily formulated in the large $N_{\rm C}$ limit, but is easily
included for $N_{\rm C}=3$, and hence brings with it some other finite
$N_{\rm C}$  effects.  The
combined effect of the careful treatment of the non-linear equations
and the configuration mixing is to give {\em substantial} central
mid-range attraction for the NN system that is in qualitative
{\em agreement} with the data.
To go from this work to a theory
that can be confronted with experiment in detail
is a difficult challenge.  There
are $R$ dependent corrections to inertial parameters to include, there
are dynamical quantum corrections, there are non-adiabatic effects that
are particularly important at small $R$, and there are other mesons to
include in the Skyrme Lagrangian. All these are under study.  The
success so far encourages us to proceed and to suggest that the results
obtained so far will be robust under these refinements.

     The success of the Born-Oppenheimer approach to the nucleon-nucleon
interaction in the Skyrme model raises interesting questions about
what ``nucleons" are when they are interacting.  We find that the
state that is two nucleons at large separation becomes, adiabatically,
a state with delta components as the nucleons approach.
For the Skyrmion-Skyrmion system itself this distortion and deformation
under interaction is even more dramatic.  What then should we say about
the interacting nucleons?  That there is a big delta component
in their wave function?  That they are distorted by the interaction?
In some sense this is true, but in the spirit of the Born-Oppenheimer
approximation, it is not.  The adiabatic potential derived from
the Born-Oppenheimer prescription involves state mixing, in fact
much of its attraction comes from that mixing, but that adiabatic
potential is to be used as an effective potential
{\it between nucleons}.  Think of the
van der Waals potential between hydrogen atoms.  It comes from
virtual excitation of the atoms, principally to the first
$L=1$ state, but the potential is to be used between atoms
in their ground state.  One usually
does not talk about the percent of
atomic $L=1$ state in the hydrogen molecule.  Following that example,
we should not talk about the percent of delta in the state
of two interacting nucleons.  On the other hand, one could take
apart the Born-Oppenheimer calculation and quote such a percent.
It is really a matter of what basis one uses.  Many discussion
of delta's in the nucleus are unclear about this
ambiguity.  Our calculation helps to emphasize the
perspective dependence of these discussions.  We plan
to return to this topic both for the nucleon-nucleon
scattering system and for the deuteron in
a subsequent paper, as well as to the question of the
interaction effects on the nucleon mass parameter.
We have also seen that  corrections to the adiabatic approximation
make it possible to compute transition potentials. These can be used to
calculate Delta  production in $NN$ collisions. In this
paper we have only scratched the surface of this interesting
problem.

     In summary we have shown that including the effects of channel
coupling ($\Delta$ mixing) in the projection of the nucleon-nucleon
interaction from recent studies of the potential in baryon number
two Skyrmion systems, substantially increases the strength of the
mid-range central attraction bringing it into qualitative agreement
with experiment.
To obtain this result we introduce some
quantum corrections (equivalently corrections for the finite  number of
colors).  Many more sophisticated quantum corrections remain
to be made.  But our results show that a non-perturbative
approach to the problem of obtaining the ``static'' nucleon-nucleon
interaction from QCD based on the Skyrme approach has great promise.
We are therefore encouraged to follow up the first success reported here.

\acknowledgments
We wish to acknowledge Atsushi Hosaka for his contributions
in the initial stages
of this work. We also would like to thank Jochem Wambach for stimulating
discussions.

This work was partially supported by the U.S. National Science Foundation.
\appendix{Details of the perturbative expansion}
We  exhibit details of the calculation for
the first matrix element, for the part of $\tilde H$ proportional to
$W$. We again make use of an equation from \cite{OBBA}, Eq.\
(30).
\begin{eqnarray}
\lefteqn{
\left<NN\right| W \left|N\Delta\right>
\left<N\Delta\right| W \left|NN\right>}\nonumber\\
& = & \left<N\right|T^\alpha_{pi} \left|N\right>
\left<N\right| T^\alpha_{qj} \left|N\right>
\left<N\right|T^\beta_{pi} \left|\Delta\right>
\left<\Delta\right| T^\beta_{qj} \left|N\right>
\nonumber\\
& = & \left<N\right|T^\alpha_{pi} \left|N\right>
\left<N\right| T^\alpha_{qj} \left|N\right>
(\left<N\right|T^\beta_{pi} T^\beta_{qj} \left|N\right>  -
      \left<N\right|T^\beta_{pi} \left|N\right>
\left<N\right| T^\beta_{qj} \left|N\right>)
\nonumber\\
& = & \left<NN\right|
\frac{N_{\rm C}^4}{9}\tau^1_p\sigma^1_i\tau^1_q\sigma_j  P_N^2
\{\mbox{$\frac{1}{2}$}Q_N^2
(\mbox{$\frac{2}{3}$}\delta_{ij}-
\mbox{$\frac{i}{3}$}\epsilon_{ijl}\sigma^2_l)
(\mbox{$\frac{2}{3}$}\delta_{pq}-
\mbox{$\frac{i}{3}$}\epsilon_{pqr}\tau  ^2_r)\}
\left|NN\right>
\nonumber\\
&=& \left<NN\right|
N_{\rm C}^4\mbox{$\frac{P_N^2Q_N^2}{18}$}
(2+\mbox{$\frac{2}{3}$}\sigma^1\!\cdot \sigma^2 )
                  (2+\mbox{$\frac{2}{3}$}\tau  ^1\!\cdot \tau  ^2 )
	\left|NN\right>.
\end{eqnarray}
Here we have used the finite-$N_{\rm C}$ correction terms $P_N$,
defined before, and
$Q_N\equiv \sqrt{(1-1/N_{\rm C})(1+5/N_{\rm C})}$.
In this manner we find that
\begin{eqnarray}
\lefteqn{\left<NN\right| \tilde H(\vec R) \left|N\Delta\right>
\left<N\Delta\right| \tilde H(\vec R) \left|NN\right>}
\nonumber\\
& = &
\left<NN\right|
P_N^2Q_N^2(2 + \mbox{$\frac{2}{3}$} \tau^1\cdot\tau^2) \{
\mbox{$\frac{1}{9}$}(v_2^2+2v_3^2)   +
\mbox{$\frac{1}{27}$}(v_2^2-v_3^2)   \sigma^1\!\cdot \sigma^2    +
\nonumber\\&&\hspace*{2cm}
\mbox{$\frac{1}{27}$}(v_3^2-v_2v_3)
(3\sigma^1\!\cdot \hat R\sigma^2 \!\cdot\hat R-\sigma^1\!\cdot\sigma^2 )
\}\left|NN\right>,
\label{eq:PerND}
\end{eqnarray}
and for an intermediate state with two $\Delta$'s,
\begin{eqnarray}
\lefteqn{
\left<NN\right| \tilde H(\vec R) \left|\Delta\Delta\right>
\left<\Delta\Delta\right| \tilde H(\vec R) \left|NN\right>}
\nonumber \\
& = &
\left<NN\right|
Q_N^4(\mbox{$\frac{4}{3}$}  - \mbox{$\frac{2}{9}$} \tau^1\cdot\tau^2) \{
\mbox{$\frac{1}{3}$}(v_2^2+2v_3^2)-
\mbox{$\frac{1}{18}$}(v_2^2-v_3^2)   \sigma^1\!\cdot \sigma^2 -
\nonumber\\&&\hspace*{1cm}
\mbox{$\frac{1}{18}$}(v_3^2-v_2v_3)
(3\sigma^1\!\cdot \hat R\sigma^2 \!\cdot\hat R-\sigma^1\!\cdot\sigma^2 )
\}\left|NN\right> .
\label{eq:PerDD}
\end{eqnarray}
When we use the fact that the energy needed for excitation of a single
$\Delta$,
which we denote by $\delta$, is half that for excitation
of two $\Delta$'s, we find
\begin{eqnarray}
V^{(2)}_{\rm exact} & = & -\frac{Q_N^2}{\delta}\{
[(\mbox{$\frac{4}{9}$}P_N^2 + \mbox{$\frac{2}{9}$}Q_N^2) +
(\mbox{$\frac{4}{27}$}P_N^2-
\mbox{$\frac{1}{27}$}Q_N^2)\tau^1\!\cdot\tau^2]
(v_2^2+2v_3^2) + \nonumber\\
&&   (\sigma^1\!\cdot \sigma^2 )
[(\mbox{$\frac{4}{27}$}P_N^2 - \mbox{$\frac{1}{27}$}Q_N^2) +
(\mbox{$\frac{4}{81}$}P_N^2+
\mbox{$\frac{1}{162}$}Q_N^2)\tau^1\!\cdot\tau^2]
(v_2^2-v_3^2)\nonumber\\
&&
(3\sigma^1\!\cdot \hat R\sigma^2 \!\cdot\hat R-\sigma^1\!\cdot\sigma^2 )
[(\mbox{$\frac{4}{27}$}P_N^2 - \mbox{$\frac{1}{27}$}Q_N^2) +
(\mbox{$\frac{4}{81}$}P^2+
\mbox{$\frac{1}{162}$}Q_N^2)\tau^1\!\cdot\tau^2]
(v_3^2-v_2v_3)\}.\nonumber\\
\end{eqnarray}
Of course this can be rewritten by introducing spin and isospin
projection operators,
\begin{eqnarray}
P^\tau_0 &=& \frac{1-\tau_1\!\cdot\tau_2}{4},\nonumber\\
P^\tau_1 &=& \frac{3+\tau_1\!\cdot\tau_2}{4}, \\
\tau^1\!\cdot\tau^2 & = & -3P^\tau_0 + P^\tau_1.\nonumber
\end{eqnarray}
We thus get the form (subscripts now denote isospin projections)
\begin{eqnarray}
V^{(2)}_{\rm exact} & = &  -\frac{Q_N^2}{\delta}\{
[\mbox{$\frac{1}{3}$}Q_N^2 P^\tau_0 + (\mbox{$\frac{16}{27}$}P_N^2+
\mbox{$\frac{5}{27}$}Q_N^2)
P^\tau_1]
(v_2^2+2v_3^2) + \nonumber\\
&&   (\sigma^1\!\cdot \sigma^2 )
[-\mbox{$\frac{1}{18}$}Q_N^2 P^\tau_0+
(\mbox{$\frac{16}{81}$}P_N^2-\mbox{$\frac{5}{162}$}Q_N^2)P^\tau_1]
(v_2^2-v_3^2)\nonumber\\
&&
(3\sigma^1\!\cdot \hat R\sigma^2 \!\cdot\hat R-\sigma^1\!\cdot\sigma^2 )
[-\mbox{$\frac{1}{18}$}Q_N^2 P^\tau_0+
(\mbox{$\frac{16}{81}$}P_N^2-\mbox{$\frac{5}{162}$}Q_N^2)P^\tau_1]
(v_3^2-v_2v_3)\}.\nonumber\\
\end{eqnarray}

\figure{The central potential $V_{\rm c}^T$ as a function of $R$ in
the region $ 1$ to $2$ fm for the $T=0$ channel.  The solid
line gives the nucleons only result of \protect{\cite{WW}}.  The longer dashed
line is the result of the state mixing in perturbation
theory and the shorter dashed line of the full Born-Oppenheimer
diagonalization.  The dotted line is the Reid soft core potential
in this channel. }

\figure{Same as Figure 1 but for $T=1$.}

\figure{The spin dependent potential $V_{\rm s}^T$ as a function of $R$ in
the region $ 1$ to $2$ fm for the $T=0$ channel.  The solid
line gives the nucleons only result of \protect{\cite{WW}}.  The longer dashed
line is the result of the state mixing in perturbation
theory and the shorter dashed line of the full Born-Oppenheimer
diagonalization.  The dotted line is the Reid soft core potential
in this channel. }

\figure{The same as Figure 3 but for the $T=1$ spin-dependent
potential. }

\figure{The tensor potential $V_{\rm t}^T$ as a function of $R$ in
the region $ 1$ to $2$ fm for the $T=0$ channel.  The solid
line gives the nucleons only result of \protect{\cite{WW}}.  The longer dashed
line is the result of the state mixing in perturbation
theory and the shorter dashed line of the full Born-Oppenheimer
diagonalization.  The dotted line is the Reid soft core potential
in this channel. }

\figure{The same as Figure 5 but for the $T=1$ tensor
potential.}

\figure{The potential in the $^3S_1$ ($T=0$) channel as a
function of $R$.  The solid line gives the nucleons only result
of \protect{\cite{WW}}.  The dashed line is the result of state mixing in
a full Born-Oppenheimer diagonalization, and the dotted line
is the Reid soft core potential in this channel.  }

\figure{ Same as Figure 7 but for the $^1S_0$ ($T=1$) channel. }

\figure{The Born-Oppenheimer energy eigenvalues coming from
diagonalization of the adiabatic energy in the $J=1^-$, $T=1$
channel as a function of the adiabatic variable $R$.
The lowest energy state goes over to the $NN$ state for large $R$.
The next three states go over into the $N\Delta$ states, and the
upper two into the $\Delta\Delta$ states.}

\figure{The percentage admixture of configurations
in the lowest ($N-N$ ) adiabatic
state of the $ J=1^-$ $T=1$ channel as a function of $R$.
The solid line is the $NN$ component, and the seecond
largest component, given by the short-dashed line is
the state $\left|N\Delta,L=1,S=2\right>$.
}

\figure{The effective mass of the ``nucleon" in the lowest
$ J=1^-$ $T=1$ channel as a function of $R$.  This is
taken to be twice the reduced mass.
}

\figure{The velocity independent transition potential from the
lowest ($NN$) adiabatic
state of the $ J=1^-$ $T=1$ channel as a function of $R$ for
transitions to each of the five other channels.
The numbers in the legend refer to the order of the
states in Fig.~9, where one is the lowest energy adiabatic state.
}

\figure{Same as Figure 12 but for the velocity dependent
transition potential.  Each term is understood to be multiplied
by $-i \beta$ where $\beta = {v}/{c}$. }

\begin{references}
\bibitem{remarkdoughnut}
It is not immediately clear whether one
can describe the symmetries in the doughnut in this way. We have studied
this problem in some detail and have been able to prove that this is
possible, but we have not yet been able to derive a sensible physical
mechanism for the partial symmetry restoration of the doughnut.
\bibitem{remarkNP}
Note the close analogy to an axially deformed nucleus, which can be
represented by a point on the sphere $S_2$. In both cases this represents
a state that is not the most general deformed body for the group under
consideration, which leads to selection rules, $K_z=0$ for the nuclear
case, and $\vec{K}=\vec{I}+\vec{S}=0$ for the case of the Skyrmion.
\end{references}
\end{document}